\documentstyle[epsfig]{mn}

\begin{document}
\title
[Spatial Filters for Optical Interferometers]
{Numerical Simulations of Pinhole and Single Mode Fibre Spatial
Filters for Optical Interferometers}
\author
[J.W. Keen, D.F. Buscher and P.J. Warner]
{J.W. Keen,\thanks{E-mail: jwk@mrao.cam.ac.uk} D.F. Buscher and P.J. Warner\\
Astrophysics Group, Cavendish Laboratory, Madingley Road,
Cambridge, CB3 0HE}
\maketitle

\begin{abstract}
We use a numerical simulation to investigate the effectiveness of pinhole
spatial filters at optical/IR interferometers and to compare them with
single-mode optical fibre spatial filters and interferometers without
spatial filters. 
We show that fringe visibility measurements in interferometers containing
spatial filters are much less affected by changing seeing conditions
than equivalent measurements without spatial filters.
This
reduces visibility
calibration uncertainties, and hence can reduce
the need for frequent observations of 
separate astronomical sources for calibration of
visibility measurements.  We also show that spatial filters can increase the
signal-to-noise ratios of visibility measurements and that pinhole
filters give
signal-to-noise ratios within 17$\%$ of values obtained with
single-mode fibres for aperture diameters up to $3r_0$.  Given the
simplicity of the use of pinhole filters we suggest that
it represents a competitive, if not optimal, technique for spatial
filtering in many current and next generation interferometers.
\end{abstract}


\section{Introduction}
Minimising measurement uncertainties in visibility observations with
optical/IR interferometers is one of the major challenges facing any
designer of a modern interferometric array.
These uncertainties arise from both instrumental and 
atmospheric effects. The instrumental effects result from aberrations
in the optical train and are usually fixed or slowly changing.
Atmospheric effects are due to 
turbulence which causes rapidly-varying phase corrugations in stellar
wavefronts.
These corrugations
corrupt the measured amplitudes and phases of the interference fringes.

The use of the closure phase allows most of the object 
phase information to be recovered,
and closure phase accuracies of a few degrees can be achieved after
several seconds of averaging on a bright source. In contrast, the
fringe amplitude or visibility 
is much harder to measure accurately. For example, the mean square
visibility of a point source, which ought
to have a constant value of 100\%, is typically observed to be less
than this value and to vary
by 10-50\% on timescales of minutes to hours.  This reduction in
fringe visibility is due to mismatches in the shapes of the two
wavefronts being interfered, caused by atmospheric and instrumental
effects.  The atmospheric mismatches vary on millisecond timescales but
even the mean effect of these mismatches taken over several seconds
varies due to changes in the quality of the seeing.

Some improvement can be obtained by observing a nearby
point source and using this to estimate the visibility
reduction. However, the use of a calibration source depends on the 
assumption that the
visibility losses remain constant over the several minutes required to
switch between calibrator and science objects.  This is a poor
assumption for atmospheric effects and as a result the calibrated visibilities
still show variations at around the 10\% level.  Furthermore, the
constant switching between science and calibration sources
dramatically reduces the usable
observing time for an interferometer, and hence the amount of science
that can be done with an instrument.  
Thus any system which stabilises the visibility losses is valuable
because it can reduce 
the reliance on a
calibration source.

A major step in stabilising visibility losses was made when it was
realised that spatially filtering the beams entering the beam
combination system
would remove the spatial phase perturbations across the incoming
wavefronts \cite{Shaklan88} and hence remove the major atmospheric
contribution to visibility loss. Initial results with this technique
have been extremely promising, reducing calibration errors on
visibility measurements by as much as two orders of magnitude
\cite{Foresto98}.
Such high-precision measurements
are required for many of the most exciting astrophysical programmes
for current and future arrays such as direct measurement of Cepheid
pulsation.  Consequently spatial filtering is being actively
pursued and several interferometer projects are now
using or designing spatial filtering systems.

Most of the work on spatial filters has been based on the use of
single-mode optical fibres.  In this paper we present a detailed
analysis of a competing approach, where spatial filtering is provided
by focusing a collimated beam onto a pinhole (see Prasad \& Loos
\shortcite{Prasad92} and St. Jacques \shortcite{StJacquesPhD}).
Despite the relative simplicity of this approach the use of pinholes
has been largely ignored by the astronomical community under the
impression that they provide inferior results.  Our analysis compares
the performance of pinholes and optical fibres under a wide range of
conditions.  We demonstrate that pinholes represent a simple and
effective method of spatial filtering. The use of pinholes instead of
optical fibres leads to little loss of performance when used with
aperture sizes typical of most current and planned interferometers.
We argue that, when the ease of implementation of
pinhole systems is taken into account, it is likely that pinhole
systems will out-perform fibre systems in practice.

In section \ref{sec:filters} we introduce the physical basis for
spatial filtering, and we develop our numerical model in section
\ref{sec:measure}.  Our simulation and analysis technique is
outlined in section \ref{sec:technique} and the results are presented
and discussed in sections \ref{sec:results} and \ref{sec:discussion}
respectively.

\section{Using Pinholes and Single mode optical fibres as spatial
filters}\label{sec:filters}

\subsection{Pinhole spatial filters}\label{sec:pinholes}

Spatial filtering with pinholes is very simple in concept. A lens or
mirror is used to focus the collimated beam from a star into the centre
of a pinhole which is comparable in size to the diffraction limit of
the aperture.  The light is transmitted through the pinhole and
projected onto another lens to produce a filtered beam which can be
interfered with a filtered beam from another telescope. As the electric field
distribution at the focus (i.e. the image plane) is simply the Fourier
Transform of wavefront across the collimated beam (i.e. the aperture
plane), the higher-spatial-frequency perturbations do not pass through
the pinhole and the hence the output wavefront is much smoother than
that of the input beam.

The visibility of the fringes
formed by interfering two such smoothed beams will be higher than it
would be for
the unfiltered beams and will depend less strongly on the strength of
the perturbations in the input wavefront.
Smaller pinholes will remove more of the high-frequency wavefront perturbations
than larger ones, albeit at the expense of the optical 
throughput;
this trade-off is investigated further in section \ref{sec:simtech}.

\subsection{Single mode fibre spatial filters}\label{sec:fibres}

We can replace the pinhole described in section \ref{sec:pinholes}
with a short length of 
single-mode fibre to produce a similar effect. The single-mode fibre
will reject light which does
not match the first guided mode of the fibre
\cite{Jeunhomme83}. This mode is approximately Gaussian shaped and is
typically of
similar size to the diffraction pattern of the aperture being used. The beam
leaving the fibre is always the same shape in amplitude and phase,
independent of the shape of the input wavefront. 
For this reason, when the beams emerging from two fibres are
interfered
there will be no reduction in visibility
due to phase corrugation, and hence single mode fibres are
often referred to as ``perfect spatial filters''.  However, there are
practical difficulties in efficiently coupling light into single-mode
fibres, and these
are discussed in section \ref{sec:discussion}.

\section{Visibility measurements and auto-calibration} \label{sec:measure}

In this paper we concentrate on the effects of spatial filters on
the measured fringe visibility and will ignore their effects
on closure phases. We define the mean square visibility as
the mean squared fringe amplitude normalised by the square of the mean
intensity of the fringe. For both pupil-plane and image-plane
beam-combination systems, it can be shown \cite{Buscher88}
that the mean square visibility
will be given by 
\begin{equation}
<V^2> = \frac{4\Bigl<\Bigl| \int\limits_{all \vec{r}} A_1(\vec{r},t)
A_2^\ast(\vec{r},t) d\vec{r} \Bigr|^2\Bigr>}{\bigl<\int\limits_{all \vec{r}} |A_1(\vec{r},t)|^2 d\vec{r} +
\int\limits_{all \vec{r}} |A_2(\vec{r},t)|^2
d\vec{r}\bigr>^2}\label{eq:vis_gen}
\end{equation}
where $A_1(\vec{r},t)$ and $A_2(\vec{r},t)$ denote the instantaneous spatial
profiles at time $t$ of the electric fields across the two beams being
interfered, $\vec{r}$ denotes a position in the aperture plane
and angle brackets denote a mean over $t$. It has been assumed that
the fringe amplitude is sampled in a time short compared to the time
taken for the wavefront shape to change (the seeing coherence time), 
but averaged over a period
which is long compared to this coherence time.

A useful property of
this definition of the visibility is that independent random 
fluctuations in the overall
intensities of the two beams being interfered has no effect on
$<V^2>$: only differences in the {\em time-averaged} beam
intensity can cause a reduction in the mean squared visibility
\cite{ShaklanPhD}.  Such differences in the average will typically be
caused by instrumental effects, and we would expect these to
vary slowly compared to
atmospheric effects.  

If the use of spatial filtering reduces the visibility loss due to
atmospheric phase perturbations to a negligible level then the
measured visibilities (calibrated for instrumental effects) can be taken
as the true visibility of the source.  This
removes the need for frequent switching to a
separate calibration source, thereby increasing the useful observing
time of the interferometer. Hereafter we will call this technique ``auto-calibration''.

\section{Technique}\label{sec:technique}
\subsection{Simulation techniques}\label{sec:simtech}

We employ a numerical simulation to investigate the
performance of the pinhole and fibre spatial filtering systems.
Our analysis considers two
apertures observing a distant, unresolved, source.  
We further assume that the light intensity incident
on each aperture is equal and that both apertures have identical optical
throughputs to a beam combiner which forms interference fringes. 
We assume the fringe visibility is
measured according to equation~\ref{eq:vis_gen}.
This
simulates observations which would result in unit visibility in the
absence of atmospheric effects.  

We have assumed that the atmospheric turbulence obeys
Kolmogorov statistics with an outer scale much larger
than the size of any aperture. We use the technique of McGlamery
\shortcite{McGlamery76} to generate random wavefront phase
perturbations whose power spectrum is given by \cite{Roddier81}
\begin{equation}
\Phi(\omega) = \frac{0.0229}{r_0^{5/3}}\omega^{-11/3}\label{eq:ps}
\end{equation}
where $r_0$ is the Fried parameter of atmospheric seeing
\cite{Fried66}.

Once two random phase screens have been generated they are used
to produce two arrays representing the complex electric field amplitudes
$A_1(\vec{r},t)$ and $A_2(\vec{r},t)$ of the two incoming
wavefronts. These wavefronts are of constant intensity, i.e. there is
no atmospheric scintillation.  These
arrays are then multiplied by a circular aperture function.

We compensate for 
the tilt component of the wavefront perturbations in a way that 
simulates the operation of a fast autoguider.
A fast quad-cell wavefront sensor measures the wavefront tilt and
adjusts it to equalise the flux in
the four quadrants of the image formed by Fourier transforming and
squaring the wavefront amplitude.
This system of fast guiding is chosen to
mimic the auto-guider used at COAST (Baldwin et al. 1994a,b) and
many other interferometers such as NPOI \cite{Hutter98} and the VLTI
\cite{Beckers90} auxiliary telescope array. 

Different amounts of delay in the tip-tilt correction system are
simulated assuming a single layer of ``frozen'' turbulence passing
at speed $v$ over each telescope. A guider with a fixed delay $\Delta t$ is
simulated by using the tilt correction signal derived from a patch of
turbulence offset by $v\Delta t$ from the patch of turbulence used to
derive the interferometric signal.    

The simulated wavefronts are then spatially filtered using
pinholes or single mode fibres. 
Pinhole spatial filters are simulated by Fourier transforming
the wavefront amplitudes to produce the image-plane electric field
distribution and then multiplying this distribution by a
circular transmission function which is unity inside the pinhole and
zero outside it. 

For the single-mode fibres we assume a step-index cylindrical core.
As the wavefront emerging from the fibre is independent of the
incoming beam we need only consider the total amount of light that is
coupled into the fibre core.  The intensity coupling
efficiency of an optical fibre is computed from \cite{Wagner82}
\begin{equation}
\Biggl| \int\limits_{\bf \tilde{S}}\tilde{\Psi}_f\tilde{\Psi}_b {\bf d\tilde{S}} \Biggr|^2
\end{equation}
where $\tilde{\Psi}_f$ is the far-field pattern of the 
the mode propagating inside the fibre and
$\tilde{\Psi}_b$ is the aperture-plane distribution of the wavefront
being focussed on to the fibre end.
The electric field
patterns ($\tilde{\Psi}_f$ and $\tilde{\Psi}_b$) have to be normalised
such that 
\begin{equation}
\int\limits_{\bf\tilde{S}} |\tilde{\Psi}_f|^2 {\bf d\tilde{S}}
=
\int\limits_{\bf\tilde{S}} |\tilde{\Psi}_b|^2 {\bf d\tilde{S}}
 = 1
\end{equation}
We make the approximation that the first guided mode of a step-index
fibre is a Gaussian \cite{Shaklan88}, so that the field
distribution in the fibre is given by
\begin{equation}
\Psi_f = \sqrt{\frac{2}{\pi}}\frac{1}{w} \exp\Biggl({-\frac{r^2}{w^2}}\Biggr)
\label{eq:mode}
\end{equation}
where $r$ is the radial distance from the centre of the fibre and $w$
is the `Gaussian width' of the fundamental mode of the fibre. This
width is related to the core size of the fibre. The far-field
pattern of the fibre is the Fourier transform of this pattern and
is therefore also Gaussian.

\subsection{Data analysis}\label{sec:data}

The visibility and signal-to-noise ratio (SNR) 
of an observation are calculated in terms of two
intermediate quantities:-
\begin{equation}
\alpha(t) = \Biggl| \int\limits_{\mbox{all} \vec{r}} A_1(\vec{r},t)
A_2^\ast(\vec{r},t) d\vec{r} \Biggr|^2
\end{equation}
and
\begin{equation}
\beta(t) =  \int\limits_{\mbox{all} \vec{r}} |A_1(\vec{r},t)|^2 d\vec{r} +
\int\limits_{all \vec{r}} |A_2(\vec{r},t)|^2 d\vec{r}
\end{equation}
where $A_1(\vec{r},t)$ and $A_2(\vec{r},t)$ represent the electric
fields across the two wavefronts at some time $t$. 
The value  $\alpha$ represents the instantaneous
fringe amplitude and $\beta$ gives the sum of the
beam intensities.
These values are averaged over a large
number (typically between 50 and 3000) 
of independent random wavefront realisations to reduce their statistical
errors. 

The visibility we measure (see equation \ref{eq:vis_gen}) is given by
\begin{equation}
V_{rms} = \sqrt{<V^2>} = \sqrt{\frac{4<\alpha(t)>}{<\beta(t)>^2}}.\label{eq:vis}
\end{equation}

At low light levels (where the fringe measurements are
photon-noise-limited and the 
SNR per sample is much less than unity) the signal-to-noise ratio of the 
mean squared visibility 
is given by \cite{Dainty79}: 
\begin{equation}
SNR \propto \frac{<\alpha(t)>}{<\beta(t)>}\label{eq:snr}
\end{equation}
This signal-to-noise ratio depends on the star brightness and the instrument
characteristics, but what we are interested in here is the relative
values of the SNR for different spatial filtering configurations and
aperture sizes. The values presented here all correspond to
measurements using the same number
of photons per $r_0$-sized patch, where $r_0$ is the Fried parameter
of the seeing. This implies constant values for $r_0$ and the
intensity of the source.

\subsection{Dimensionless units}\label{sec:units}

In order to make the results of this analysis applicable to a variety
of wavelengths, seeing conditions and aperture sizes we express them
in terms of two 
dimensionless units of distance and one of time.  In the
aperture plane, the aperture sizes are stated as multiples of the Fried
parameter, $r_0$.

In the image plane of a lens of focal length $f$,
the dimensionless unit of length is related to an actual distance,
$x$, by  $\displaystyle\frac{xD}{f\lambda}$
where $D$ is the beam diameter and $\lambda$ is the wavelength
of the light.  Using this system the size of the diffraction limited
focal spot is independent of primary beam size and wavelength: the
radius of the first Airy dark ring will always be 1.22 
dimensionless units.

Guider delays are measured in terms of the atmospheric coherence time
$t_0=0.31r_0/v$ where $v$ is the effective wind speed.

\section{Results}\label{sec:results}

We present our results for both pinhole and single-mode fibre filters
and compare them to results for telescopes without spatial filtering.  We consider the
effect of filtering on the signal-to-noise ratio of visibility
observations and then analyse the visibility loss associated with a
pinhole filter.

\subsection{Signal-to-noise ratios}\label{sec:snr}

\begin{figure}
\begin{center}
\psfig{figure=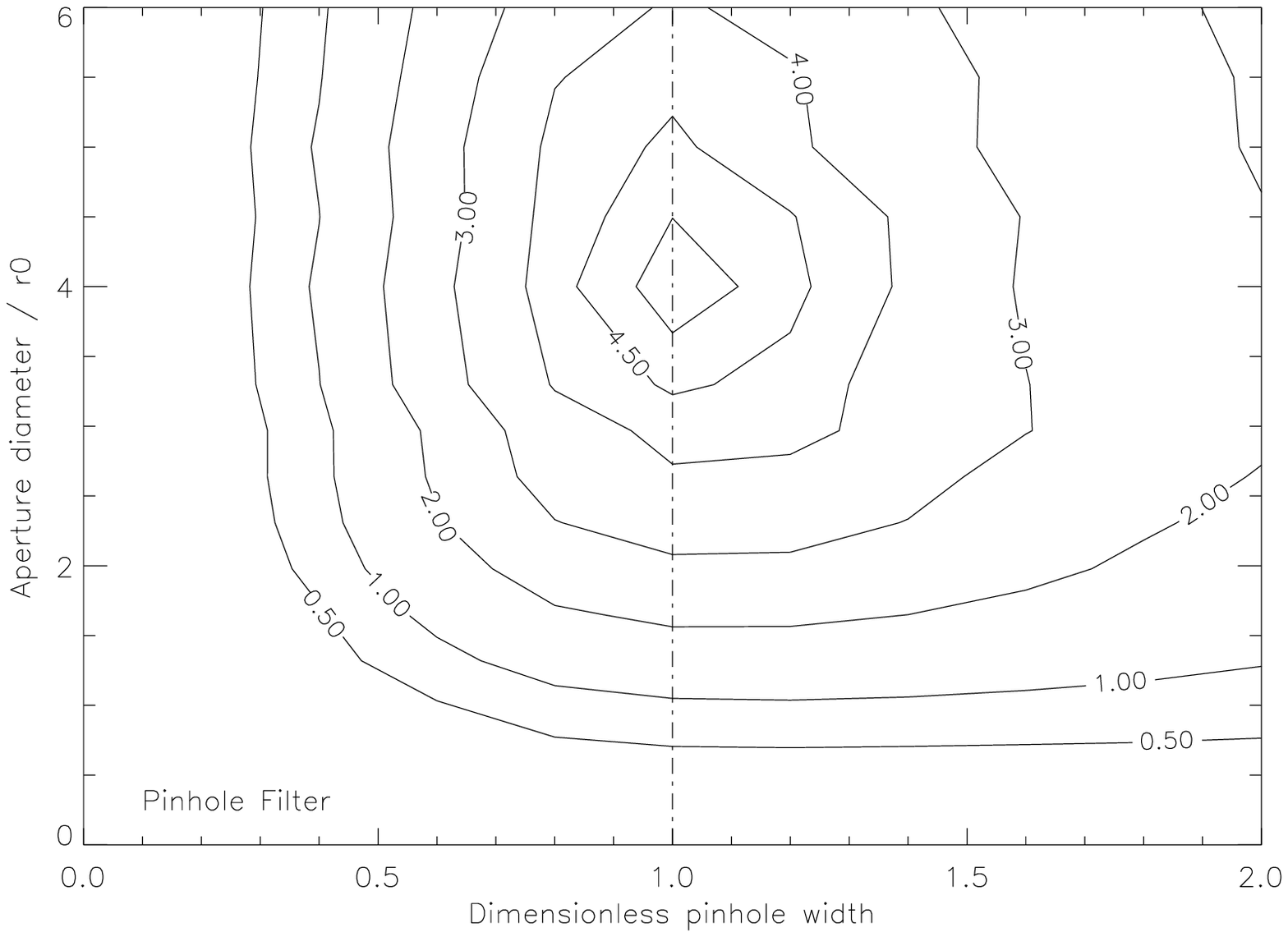, width=8cm}\ \\
\psfig{figure=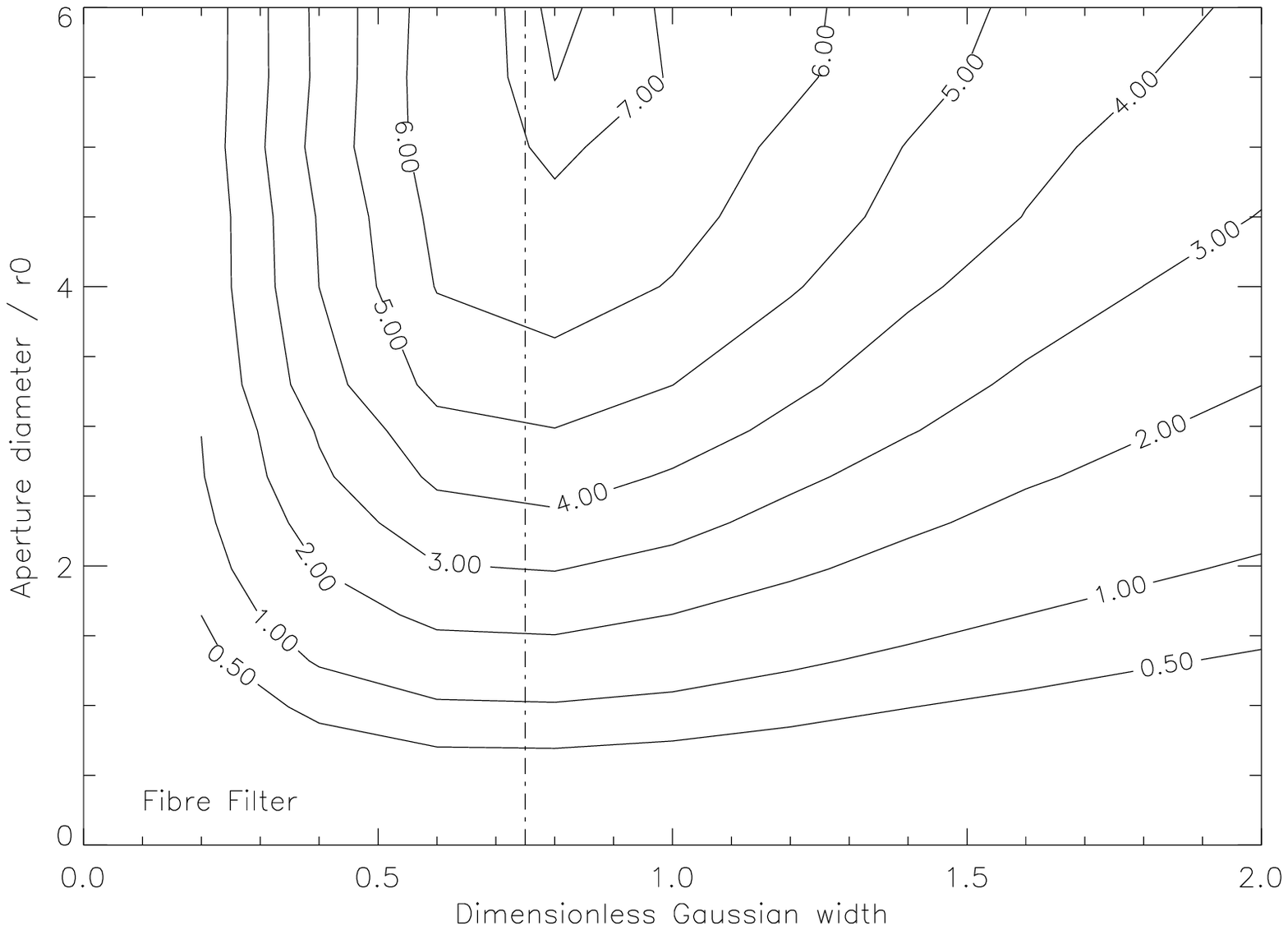, width=8cm}
\caption{Signal-to-noise ratio for pinhole (upper plot) and fibre (lower plot) filters
of various filter and aperture sizes.  A constant value of the Fried
parameter, $r_0$, is assumed.  All sizes are given in dimensionless
units (see section \ref{sec:units}).  The highest contours are 4.75 and 7.25 for
pinholes and fibres respectively.  Vertical lines show the values of
filter size chosen for all further investigations.}\label{fig:contour}
\end{center}
\end{figure}

Figure \ref{fig:contour}
shows the signal-to-noise ratio of visibility measurements for a range
of pinhole diameters and fibre core radii, assuming no guider delay. 
By finding the maximum values on these plots we
can optimize the filter and aperture size of a telescope for use with
either spatial filtering system.  For any given aperture size we can find an
optimum diameter of pinhole or fibre core radius
for which the signal-to-noise ratio is maximised.  By inspection we can
see that this optimum size varies little with aperture size
and so for reasons of clarity we shall henceforth use
a single spatial filter size for all aperture sizes.   We chose to use
pinhole filters with a radius of 1.0 dimensionless unit (see section
\ref{sec:units}) and fibres with Gaussian width (see section
\ref{sec:simtech}) of 0.75 units for subsequent calculations.  These
values are shown on figure \ref{fig:contour} as vertical lines.
For the fibre filter, the absolute peak value of the SNR lies just off the line
but we have chosen the line to be optimal for smaller apertures.

\begin{figure}
\psfig{figure=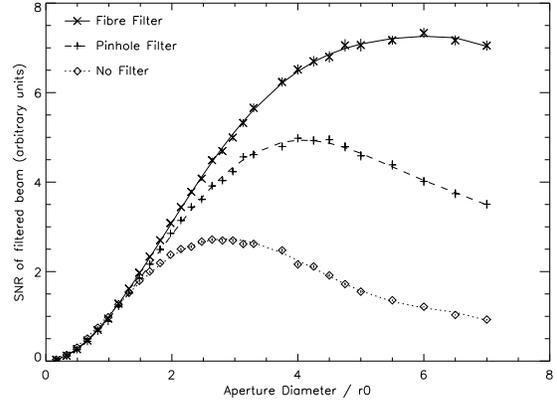, width=8cm}
\caption{The signal-to-noise ratio of a mean squared visibility
measurement using
optimized fibre and pinhole filters (see section \ref{sec:snr}).
Results are shown for circular apertures of various
diameters with perfect guiding (see section \ref{sec:simtech}).}\label{fig:snr_main}
\end{figure}

Figure \ref{fig:snr_main} shows a 1-d section taken along the marked lines in
figure~\ref{fig:contour}, along with a
comparison with the SNR for beams with no spatial filtering.  Clearly
there is an advantage in using spatial filtering to increase the signal-to-noise ratio for
any set of apertures larger than 1.4$r_0$ in diameter.  The
performance of the two spatial filtering systems is similar for
aperture diameters up to $3r_0$. At this point the
difference in SNR is less than 17$\%$.

The figure also confirms the earlier
results that the optimal signal-to-noise ratio is achieved for aperture diameters of 
2.8$r_0$ \cite{Buscher88} and 6$r_0$
\cite{Buscher94} for unfiltered and fibre filtered beams respectively,
and demonstrates that the optimal SNR for pinhole filters is reached
at around $4r_0$. This assumes a fixed photon rate per $r_0$-sized patch.

We have repeated this analysis for several values of guider delay and found
that, while increasing the guider delay does reduce the signal-to-noise ratio,
both filtering systems suffer a similar level of
SNR reduction as for an unfiltered telescope.

\subsection{Visibility measurements and visibility loss}\label{sec:visres}

\begin{figure}
\psfig{figure=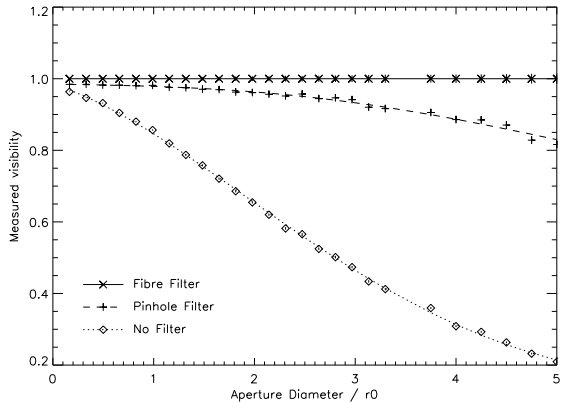, width=8cm}
\caption{The root mean squared visibility measured for a point source
at various aperture sizes. Results for interferometers incorporating
fibre spatial filters, pinhole spatial filters and no spatial filter
are presented.  
}
\label{fig:vis_unf}
\end{figure}

We would expect that there would be no visibility loss for
interferometers using fibre spatial filters because fibres are perfect
spatial filters.  Pinholes are not perfect spatial filters,
consequently there will be some residual phase corrugation for any
finite sized pinhole.  This will result in a reduction of the measured
visibility. Figure \ref{fig:vis_unf} shows the visibility of a point
source measured with a fibre spatial filter, a pinhole spatial filter
and no spatial filter for a range of normalised aperture diameters.
For the unfiltered system, the degree of visibility loss depends
strongly on the ratio of the beam size to the seeing coherence
scale. In contrast pinholes filters show significantly less variation
in fringe visibility and the visibility measured with fibre based
filters is unaffected by seeing.

\begin{figure}
\psfig{figure=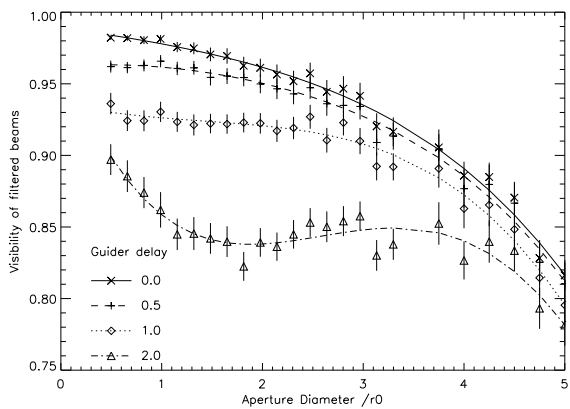, width=8cm}
\caption{The root mean squared visibility measured for a point source
at various aperture sizes and guider delays using a pinhole filter.
Guider delays are given in multiples of the coherence time, $t_0$.
Error bars are given as standard errors due to statistical
uncertainties.}\label{fig:vis_fil}
\end{figure}

The larger these variations in visibility measurement are for a given
change in seeing conditions, the more frequently we will have to
switch between sources.  If the aperture diameter $D$ remains fixed
and the seeing scale $r_0$ changes so that $D/r_0$ changes from 2 to
3, the measured visibility on a point source for an unfiltered system
changes from 0.65 to 0.45, i.e. a fractional change of 30\%. Over the
same range of seeing conditions the visibility measured with a pinhole
filter will vary by 2.5\%, in fact over the range $r_0 \leq D \leq
3r_0$ there is a variation of less than $4\%$. This range is
comparable to the largest typical variation in seeing during one
night and hence suggests that we may only need to make calibration
observations once or twice per night, i.e. auto-calibration.

Figure \ref{fig:vis_fil} shows the visibility
of a point source measured using a pinhole filtering system for a
range of seeing conditions and guider delays (note the large difference in
vertical range between figures \ref{fig:vis_unf} and
\ref{fig:vis_fil}). With a fixed guider delay, if $t_0$ changes there
can be a greater 
reduction in measured visibility than is caused by a change in the
value of $r_0$. However it is usually possible to measure the value of
$t_0$ from the interferometric data \cite{BurnsPhD} 
and therefore allow for this change.
This means that pinhole filters can be used for auto-calibration for
the majority of seeing conditions.

\begin{figure}
\begin{center}
\psfig{figure=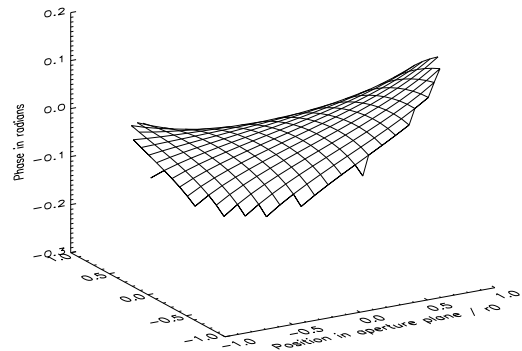, width=8cm}\ \\
\psfig{figure=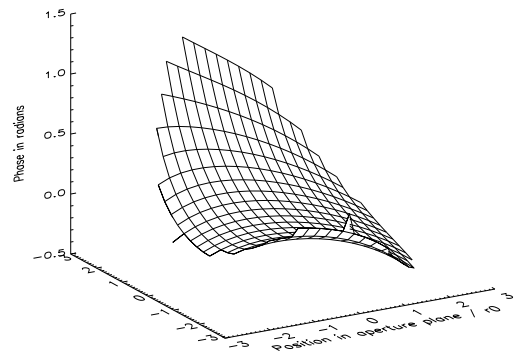, width=8cm}
\caption{Typical wavefronts from a circular aperture after passing through
a pinhole spatial filter.  The upper plot shows a wavefront produced with an
aperture diameter of $2r_0$ whereas the lower plot was produced with an
aperture diameter of $5r_0$.}\label{fig:wavefronts}
\end{center}
\end{figure}

An interesting point to consider is why the visibility measurements
drop off for pinhole filters at aperture sizes greater than $4r_0$.
We might naively believe that the tip-tilt correction system
compensates for the low spatial frequency wavefront errors and the
pinhole removes the high spatial frequency components and therefore the
visibility should be independent of seeing for a fixed pinhole size.
This is shown to be a poor assumption by inspecting typical wavefronts
leaving our simulated filter. Figure~\ref{fig:wavefronts} shows two
such sample wavefronts which have significant tilts that are larger
for poorer seeing.  This is because the tilt which the guider removes
contains contributions from all spatial frequencies and hence is not
the same as the tilt for the filtered beam. This leaves a residual
tilt in the unfiltered beams which causes a reduction in the measured
visibility. This problem could be overcome by a second level of
auto-guiding but this would probably involve a reduction in overall optical
throughput.

\section{Discussion}\label{sec:discussion}

Our results show that both pinhole and fibre spatial filtering systems
dramatically reduce the amount by which the visibility changes with a
given change in seeing. For the seeing conditions given in
section~\ref{sec:visres}, the resistance of the visibility
measurements to fluctuations in the level
of the seeing is increased by a factor of more than 10 when a pinhole
spatial filter is used.
Thus it is practical to consider making significantly fewer 
measurements of a calibration source, perhaps only a few times per night.

The visibilities measured in a system incorporating a fibre spatial
filter are in theory completely immune to changes in the seeing. This
might be interpreted as meaning that no calibration is required.
However, in practice visibility errors would be introduced by any
asymmetry in the mean coupling of light from the two arms of the
interferometer into the fibres. Such asymmetries could be introduced,
for instance if the effectiveness of the autoguiders in correcting the
wavefront tilt was different at the two telescopes.  Other effects
such as the fringes moving during the integration time over which they
are measured could also be expected to cause visibility changes at the
few percent level. Thus we would still expect some calibration to be
necessary with fibre spatial filters.

We have shown that pinhole filters can improve the signal-to-noise
ratio of visibility measurements in optical interferometers compared
to systems with no spatial filtering. For a fixed photon rate per
$r_0$-sized patch, a factor of 2 improvement in SNR is achievable with
pinholes.  For $D \leq 3r_0$ there is little difference between the
performance of pinhole and fibre filter systems, but fibre filters can
give a significantly higher signal-to-noise ratio for $D>3r_0$; the
highest SNR for a fibre filter is nearly 50\% greater than that for a
pinhole filter.

However, we must also take into account the relative ease with which
the signal-to-noise ratios computed in these simulations can be
achieved in practice. One practical consideration is that the 50\%
greater signal-to-noise ratio for fibre spatial filters is only
achieved by using aperture sizes of around $5r_0$ or greater.  By
comparison most actual interferometric telescopes typically operate
with diameters much smaller than this. For example the 1.8m outrigger
telescopes at VLTI or Keck represent only $2.4r_0$ at K band in
arcsecond conditions. At this scale we expect the signal-to-noise
ratios obtained with an ideal fibre system to be less than 10\% higher
than with an ideal pinhole system.

Another consideration is the relative ease of achieving
signal-to-noise ratios which are close to those predicted from these
numerical simulations. In practice, it is rare to achieve more than
75$\%$ of the theoretical coupling efficiency for a single-mode fibre
at optical wavelengths. There are a number of reasons for this. The
most straightforward of these is the presence of reflection losses at
the two air-glass interfaces involved in a fibre spatial filter. This
leads to an 8\% loss in light. This loss can be reduced if
anti-reflection coatings are deposited on the ends of the fibres, but
this is rare in practice. The problem of reflection is exacerbated by
errors in accurately cleaving and polishing the fibre ends which lead
to extra scattering and wavefront errors.

A more significant reason for excess light losses in practical fibre
coupling setups is the fact that any wavefront errors introduced by
the lens or mirror focusing the light into the fibre will cause a
decrease in coupling. For small wavefront errors, the coupling loss
will increase as $\sigma_\phi^2$ where $\sigma_\phi$ is the rms
wavefront error in radians. Therefore to get less than 20\% coupling
loss requires rms wavefront errors of less than $\lambda/14$.  Most
single-mode fibres are efficiently coupled with beams with an f-ratio
of about 5 \cite{Shaklan88}. This is a relatively fast focal ratio and
so to achieve diffraction-limited focusing of a beam onto a fibre
typically requires either multi-lens systems or fast off-axis
paraboloid mirrors. Manufacturing and aligning these systems to meet a
$\lambda/14$ wavefront error specification represents a formidable
challenge in the optical regime. Rohloff and Leinert
\shortcite{Rohloff91} showed that the typical throughput of a
single-mode fibre spatial filter working at optical wavelengths was
around 47\% in laboratory conditions, whereas our analysis has assumed
the theoretical maximum of 78\%. Coud{\'e} du Foresto et al
\shortcite{Foresto98} describe a total optical throughput of the FLOUR
fibre instrument, which works in the K band, as around 5-10\% with
good seeing with 45cm apertures. This can be compared to an expected
throughput of 44\% for perfect optics, assuming arcsecond seeing and
60\% detector quantum efficiency.

A similar wavefront tolerance to that of fibre systems is required for
efficient ``coupling'' of light into pinholes, but in the case of
pinholes the f-ratio of the beam is constrained only by the size of
the pinhole. By choosing a relatively large pinhole, a system with a
much larger f-ratio can be used, and this relaxes most of the optical
tolerances in the system. The pinhole itself is merely the absence of
an opaque material and cannot introduce any further aberrations. So
one would expect to approach the theoretical coupling efficiency much
more closely with a pinhole spatial filtering system than with a fibre
system, and with much simpler optics.

There are several other practical implications of implementing spatial
filtering that should be considered by any future designer
deciding which system to use. One consideration is that fibre
based systems can be incorporated into a beam transport device to
guide the light to a convenient position, reducing the number of
reflections required at an interferometer. Conversely if the spatial
filtering system is to be used as an occasional instrument and not
permanently installed then removal of a pinhole filter simply requires
moving the pinhole to one side, whereas a fibre has a finite length
and consequently requires the entire system to be removed or
recalibrated. This problem is exacerbated if the fibre is used for
beam transport.

\section{Conclusions}

We have investigated the use of pinhole spatial filters on optical/IR
interferometers to reduce the effects of atmospheric phase
perturbations on visibility measurements.  We have compared the predicted
performance of pinhole filters with single-mode fibre filter and
unfiltered interferometers.

Our major conclusions are:

\begin{enumerate}

\item Both fibre and pinhole spatial filter systems greatly reduce the
effect of changing seeing conditions on visibility measurements,
compared to an interferometer without spatial filtering.

\item Both fibre and pinhole spatial filter systems can be used to
provide ``auto-calibration'', where visibility measurements can be
made without the need to make regular observations of a separate
calibration source.

\item Pinhole spatial filters can improve the signal-to-noise ratio of visibility
measurements on optical interferometers for aperture diameters larger
than 1.4$r_0$.

\item For aperture diameters less than about $3r_0$ there is little
difference between the signal-to-noise ratio performance of pinhole and fibre spatial 
filter systems.

\item Fibre spatial filters give superior theoretical signal-to-noise
ratios for aperture diameters greater than $3r_0$. However, the extra
complexity and coupling losses involved with fibres means that
pinhole filters may be a superior solution for many current and
future optical interferometers.

\end{enumerate}

In summary any current interferometer could benefit from a pinhole
spatial filtering system as a simple but effective way to increase
overall performance and reduce measurement uncertainties.

\section{Acknowledgements}

We would like to thank all of the members of the COAST team for their
continued support and input.  Special thanks go to Chris Haniff and
John Baldwin, and to J. Jeans for illuminating contributions.

\end{document}